\documentclass[chicago]{emulateapj}
\usepackage{amsmath}

\slugcomment{Submit2 --- 19Feb2013}
\shortauthors{}

\begin{document}

\title{Gas kinematics and the Dragged Magnetic Field in the High-mass Molecular Outflow Source G192.16$-$3.84: An SMA View}

\author{Hauyu Baobab Liu\altaffilmark{1}} \author{Keping Qiu\altaffilmark{2}} \author{Qizhou Zhang\altaffilmark{3}}  \author{Josep M. Girart\altaffilmark{4}} \author{Paul T. P. Ho\altaffilmark{1,3}}

\affil{$^{1}$Academia Sinica Institute of Astronomy and Astrophysics, P.O. Box 23-141, Taipei, 106 Taiwan}

\affil{$^{2}$School of Astronomy and Space Science, Nanjing University, Nanjing 210093, China}

\affil{$^{3}$Harvard-Smithsonian Center for Astrophysics, 60 Garden Street, Cambridge, MA 02138}

\affil{$^{4}$ Institut de Ci\`{e}ncies de l'Espai, (CSIC-IEEC),Campus UAB, Facultat de Ci\`encies, C5p 2, 08193 Bellaterra, Catalonia, Spain}



\begin{abstract}
We report the Submillimeter Array (SMA) observations of the polarized 0.88\,mm thermal dust emission and various molecular line transitions toward the early B-type ($L_{*}\sim$2$\times$10$^{3}L_{\odot}$) star-forming region G192.16$-$3.84 (IRAS 05553+1631).
The peak of the continuum Stokes-I emission coincides with a hot rotating disk/envelope (SO$_{2}$ rotational temperature T$_{rot}^{SO_{2}}$$\sim$84$^{+18}_{-13}$\,K), with a north-south velocity gradient. 
Joint analysis of the rotation curve traced by HCO$^{+}$ 4-3 and SO$_{2}$ 19$_{1,19}$-18$_{0,18}$ suggests that the dense molecular gas is undergoing a spinning-up rotation, marginally bound by the gravitational force of an enclosed mass $M_{*+gas+dust}\sim$11.2-25.2\,$M_{\odot}$.
Perpendicular to the rotational plane a $\gtrsim$100/$\cos(i)$\,km\,s$^{-1}$ ($i\sim$63$^{\circ}$) high velocity molecular jet, and the $\sim$15-20\,km\,s$^{-1}$ expanding biconical cavity were revealed in the CO 3-2 emission.
The polarization percentage of the 0.88\,mm continuum emission decreases toward the central rotating disk/envelope.
The polarization angle in the inner $\sim$2$''$ (0.015\,pc) disk/envelope is perpendicular to the plane of the rotation.
The magnetic field lines, which are predominantly in the toroidal direction along the disk plane, are likely to be dragged by the gravitationally accelerated rotation.
\end{abstract}







\clearpage
\section{Introduction }
\label{chap_introduction}
The ultracompact (UC) H\textsc{ii} region G192.16$-$3.84 (c.f. Molinari et al. 1996) is a well studied B3 star-forming region at a distance of 1.52 kpc (Shiozaki et al. 2011)\footnote{Most literature assumed a distance of 2 kpc. Throughout this paper, we will update the quoted physical quantities according to the water maser parallax distance reported by Shiozaki et al. (2011).}. 
The associated 2.1\,$\mu$m point source (Indebetouw et al. 2003) was found to be embedded in a dense molecular core, which contains 290\,$M_{\odot}$ within a 0.11\,pc radius (Shepherd \& Kurtz 1999).
This source emanates an 50\,$M_{\odot}$ bipolar CO outflow extending 2.5$'$ (1.1 pc) east and 1.5$'$ (0.66 pc) west (Snell et al. 1990; Shepherd \& Churchwell 1996; Shepherd et al. 1998), which created the biconical cavity (Hodapp 1994; Shepherd et al. 1998; Indebetouw et al. 2003). 
The bipolar outflow can be further traced by H$\alpha$, [S \textsc{ii}], and 4.5 $\mu$m emission knots to $\pm$4 pc away (Devine et al. 1999; Qiu et al. 2008).

Interferometric observations of the centimeter and the millimeter continuum emission toward the 2.1\,$\mu$m source suggest the existence of circumstellar gas and dust with a total mass of 4-18 $M_{\odot}$ within a 2$''$$\times$1$''$ (3040 AU$\times$1520 AU) region (Shepherd \& Kurtz 1999; Shepherd et al. 2001; Shiozaki et al. 2011).
At the $<\sim$700 AU scale, the VLBA and VLBI (JVN and VERA) observations of the 22\,GHz H$_{2}$O maser suggested the Keplerian motion of the dense gas, and the outward motion of the bipolar outflow (Shepherd et al. 2004; Imai et al. 2006; Shiozaki et al. 2011).

This source was selected for the SMA observations of dust polarization because it represents one of the clearest cases of a massive disk/outflow system similar to the low-mass star formation (Shu et al.1987; see also Zhang et al. 1998, Cesaroni et al. 2005, Sridharan et al. 2005, Su et al. 2007, Keto \& Zhang 2010, and references therein for another massive case IRAS 20126+4104 at D=1.7 kpc, $M_{*}$=7-10\,$M_{\odot}$).
In addition, its bright thermal continuum flux at the 0.85\,mm wavelength band (2.1$\pm$0.63\,Jy in the central 15$''$ area; Shepherd et al. 2004) fulfills the required signal-to-noise ratio.
Observations of the dust polarization vectors will provide the complementary aspect of the relative importance of the magnetic field strength. 

The observing parameters for the target G192.16$-$3.84 are introduced in Section \ref{chap_obs}.
The observational results for G192.16$-$3.84 are presented in Section \ref{chap_result}.
A brief discussion is provided in Section \ref{chap_summary}.

\section{Observations and Data Reduction} 
\label{chap_obs}
We performed observations in the 0.88 mm wavelength band in the single receiver polarization mode using the SMA\footnote{The Submillimeter Array is a joint project between the Smithsonian Astrophysical Observatory and the Academia Sinica Institute of Astronomy and Astrophysics, and is funded by the Smithsonian Institution and the Academia Sinica (Ho, Moran, \& Lo 2004).} (Ho, Moran, \& Lo 2004; Marrone 2006).
The phase referencing and pointing center of the observations is R.A.=5$^{\mbox{\scriptsize{h}}}$58$^{\mbox{\scriptsize{m}}}$13$^{\mbox{\scriptsize{s}}}$.549, Decl.= 16$^{\circ}$31$'$58$''$.30 (J2000).
The primary beam size of these observations was 36$''$.
The observations tracked the frequency of 345.796 GHz at window 22 in the upper sideband. 
The spectral channel spacing was 0.7 km\,s$^{-1}$.
More details about the observations are summarized in Table \ref{table_tracks}.
The detected spectral lines in these observations are summarized in Table \ref{table_line}.

The absolute flux, passband, and gain calibrations were carried out using the MIR IDL software package. 
The typical SMA observations may be subjected to up to 15\% of uncertainty in absolute flux scales.
The calibrations of the polarization leakage (i.e. the D-term calibration) were carried out using the MIRIAD software package.
We averaged the line-free channels in the lower and upper 4GHz side bands to generate the continuum channels, and imaged the continuum emission jointly.
We carried out the imaging of the Stokes-Q and the Stokes-U components using MIRIAD, and carried out the imaging of the spectral line emission and the Stokes-I component of the continuum emission using the CASA software package.

The minimum and maximum baselines of our observations are $\sim$8 $k\lambda$ and $\sim$260 $k\lambda$, respectively. 
Our SMA observations were therefore sensitive to a maximally detectable scale of $\sim$15$''$.8 (Wilner \& Welch 1994).
Image synthesis using the Briggs Robust 0 weighting for the data collected from these observations yields a synthesized beam of 1$''$.3$\times$0$''$.87 (P.A. = 90$^{\circ}$), and a root-mean-squares (RMS) noise level of 110 mJy\,beam$^{-1}$ (0.96 K) in each 0.7 km\,s$^{-1}$ velocity channel; imaging using the naturally weighted visibilities yields a synthesized beam of 1$''$.9$\times$1$''$.4  (P.A. = 96$^{\circ}$), and a RMS noise level of 78 mJy\,beam$^{-1}$ (0.30 K) in each 0.7 km\,s$^{-1}$ velocity channel.

\begin{table}[h]\scriptsize
\begin{center}
\caption{\footnotesize{Summary of the SMA observations.}}
\label{table_tracks}
\hspace{-0.6cm}
\begin{tabular}{  | p{4cm} | p{1.32cm}p{1.32cm}p{1.32cm} | }\hline
Observing Dates  		&  2011Nov10  	& 	2012Jan06  &  2012Feb02  	\\\hline
Array Configuration	&  compact  &  subcompact  &  extended   		\\
$\tau_{\mbox{\tiny{225 GHz}}} $  & 0.05 &  0.06  & 0.05  \\
Number of Antennas &  8  &  6  &  7  \\
Time on Target Loop (hr) &      3.5    & 3.5   & 3  \\
Flux Calibrator 	& Titan			&	Ganymede			&	Callisto		 \\
Passband Calibrator	& Uranus/3C84	&	3C279		&	3C279		\\
Gain Calibrator 	& 0530+135	& 0530+135	&	0530+135	\\
Polarization Calibrator 	& 3C84	& 3C279	&	3C279	\\
Stokes-Q,U RMS (mJy\,beam$^{-1}$)	&	2	&	3	&	 2 \\\hline
\end{tabular}
\end{center}
\footnotesize{Notes. The listed RMS noise levels of the Stokes-Q and Stokes-U images are estimated based on natural weighting images. The Stokes-I image is limited by the dynamic range, and has an higher noise level (see Figure \ref{fig_pol}).}
\end{table}
\normalsize{}

\begin{table}[h]
\scriptsize{
{\renewcommand\baselinestretch{0.9}\selectfont
\scriptsize{
\begin{center}
\caption{\footnotesize{Table of the detected molecular lines.
}}
\label{table_line}
\begin{tabular}{| ll  | rcc | c |}\hline
Species & Transition & Frequency  &   E$_{u}$/k & Log$_{10}$A$_{ij}$  & Note\\
			  & &		(GHz)			&  (K)				& (s$^{-1}$) 			&	\\\hline
SO$_{2}$		&	4$_{3,1}$-3$_{2,2}$		&	332.50524		&		31		&	-3.48		&\\						
				&	8$_{2,6}$-7$_{1,7}$		&	334.67335		&		43		&	-3.90		&\\						
				&  13$_{2,12}$-12$_{1,11}$		&  345.33854		&		93			&	-3.62 & blended \\   
				&	19$_{1,19}$-18$_{0,18}$		&	346.65217		&		168		&	-3.28	&\\				
				&	16$_{4,12}$-16$_{3,13}$		&	346.52388		&		164		&	-3.47	& blended\\
SO				&	9$_{8}$-8$_{7}$					&	346.52848		&		79			&	-3.27	& blended\\
				&	8$_{8}$-7$_{7}$					&	344.31061		&		87			&	-3.28	&\\
CO			&	3-2									&	345.79599		&		33			&	-5.69&\\
CH$_{3}$OH	&	7$_{1,7}$-6$_{1,6}$++			&	335.58200		&		79			&	-3.79&\\
H$^{13}$CN	&	4-3									&	345.33976		&		41			&	-2.69& blended\\
H$^{13}$CO$^{+}$	&	4-3							&	346.99834		&		42			&	-2.49	&\\\hline
\end{tabular}
\end{center}
}
\footnotesize{\vspace{0.2cm}Notes. We omitted analyzing the blended SO$_{2}$ 16$_{4,12}$-16$_{3,13}$ and SO 9$_{8}$-8$_{7}$ lines.
}
}
}
\end{table}
\normalsize{}

\section{Results}
\label{chap_result}

\begin{figure}[h]
\hspace{0cm}
\rotatebox{-90}{
\includegraphics[width=6cm]{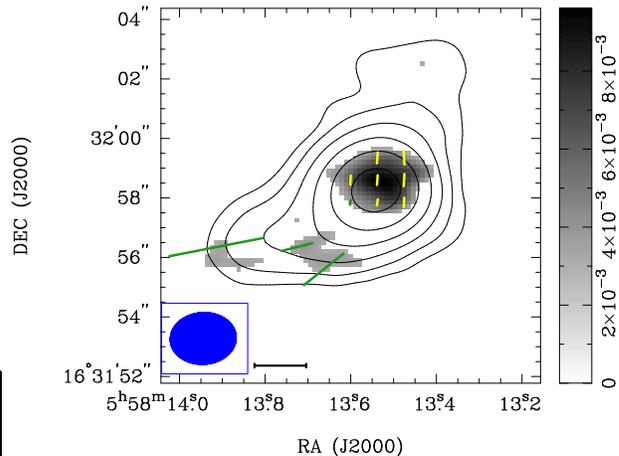}}
\caption{\footnotesize{
The naturally weighted SMA 0.88\,mm continuum image ($\theta_{\mbox{\scriptsize{maj}}}\times\theta_{\mbox{\scriptsize{min}}}$=2$''$.22$\times$1$''$.74). 
Contours show the Stokes-I image.
Contour levels are 12\,mJy\,beam$^{-1}$(3$\sigma$)$\times$[1, 2, 4, 8, 16, 32].
The polarization intensity is presented in grayscale.  
Grey-scale bar is in Jy\,beam$^{-1}$ units.
The synthesized beam of the Stokes-I,Q,U observations is shown in the lower left.
We rotate the polarization vectors by 90$^{\circ}$ and overplot them at the positions with $>$2$\sigma$ detection of the polarization intensity (green: between 2$\sigma$ and 3$\sigma$; yellow: $>$3$\sigma$; $\sigma$=1.4 mJy\,beam$^{-1}$).
With 2-3$\sigma$ detections of the polarized emission, assuming the noise characteristic is not largely deviated from the thermal noise, the 1$\sigma$ uncertainty $\sigma_{\chi}$ of the polarization position angle $\chi$ is $\sim$11$^{\circ}$-7$^{\circ}$.
The lengths of the polarization vectors are proportional to the polarization fraction.
The scale bar in the lower left represents the 10\% polarization fraction.
}}
\label{fig_pol}
\end{figure}

\subsection{Geometry}
\label{sub_geometry}
\begin{figure}[h]
\includegraphics[width=10cm]{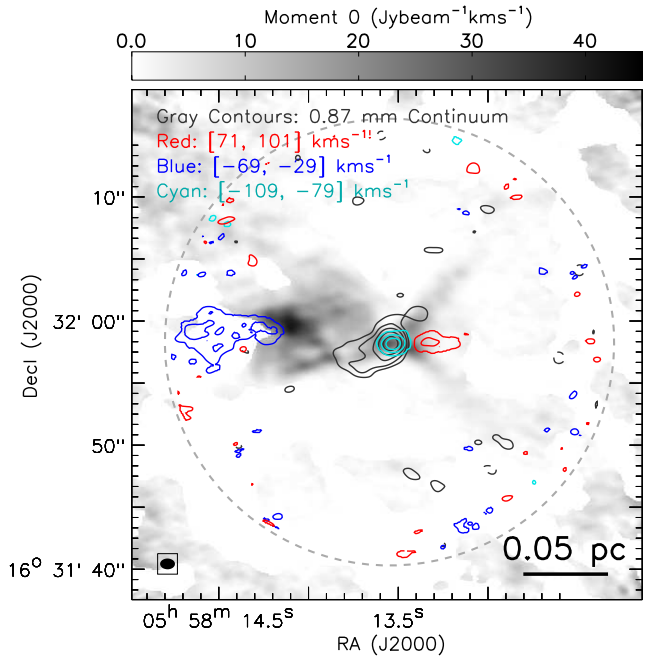}
\caption{\footnotesize{
The velocity integrated intensity map (i.e. moment 0 maps) of CO 3-2, overlaid with the 0.88\,mm continuum image (robust 0 weighted).
Dark contours show the robust 0 weighted 0.88\,mm continuum image ($\theta_{\mbox{\scriptsize{maj}}}\times\theta_{\mbox{\scriptsize{min}}}$=1$''$.3$\times$0$''$.86), with the levels of 14 mJy\,beam$^{-1}$(3$\sigma$) $\times$[-1, 1, 2, 4, 8, 16, 32].
Cyan, blue, and red contours show the integrated CO 3-2 emission in the labeled velocity ranges  (naturally weighted; $\theta_{\mbox{\scriptsize{maj}}}\times\theta_{\mbox{\scriptsize{min}}}$=1$''$.8$\times$1$''$.4).
The cyan contour levels are 2.0 Jy\,beam$^{-1}$\,km\,s$^{-1}$ $\times$[1, 2, 3]; the blue contour levels are 1.4 Jy\,beam$^{-1}$\,km\,s$^{-1}$ $\times$[1, 2].
; the red contour levels are 1.2 Jy\,beam$^{-1}$\,km\,s$^{-1}$ $\times$[1, 2].
The systemic velocity is $\sim$5.8 km\,s$^{-1}$ (see Section \ref{sub_kinematics}).
The dashed circle indicates the 18$''$ radius of the primary beam.
Contours are plotted only in the region encircled by the dashed circle.
We plot the spectral profile of the high velocity CO 3-2 components in Figure \ref{fig_spectra}.
The velocity integrated intensity map of CO 3-2  (robust 0 weighted; $\theta_{\mbox{\scriptsize{maj}}}\times\theta_{\mbox{\scriptsize{min}}}$=1$''$.2$\times$0$''$.86) in the intermediate velocity range ([-19, 1]\,km\,s$^{-1}$ and [11, 41] \,km\,s$^{-1}$) is shown by grayscale, while the channel maps in this velocity range is presented in Figure \ref{fig_chan}.
}}
\label{fig_jet}
\end{figure}

\subsubsection{The 0.88 mm Continuum Emission}
\label{subsub_ch0}
The SMA Stokes-I image (Figure \ref{fig_pol}) recovers 1.2\,Jy of flux in the central 15$''$ area. 
The previous JCMT measurement of 2.1$\pm$0.63\,Jy in the central 15$''$ area (Shepherd et al. 2004) indicates that the SMA Stokes-I image is subjected to 43$^{+13}_{-25}$\% of missing flux (see also Section \ref{chap_obs}).
The averaged intensity of the missing flux over the central 15$''$ area is 1.3-7.4 mJy\,beam$^{-1}$, which is lower than 62\% of the first contour level in Figures \ref{fig_pol} and \ref{fig_jet}.
We therefore argue that the observed geometry from the 0.88 mm continuum Stokes-I emission is only minimally affected by the missing flux issue.
According to the SED reported by Shepherd et al. (2004), the 0.88\,mm continuum emission is mainly contributed by the dust thermal emission, as well as the free-free emission from the ionized gas surrounding the central star.
The $\lesssim$1$''$ scale ultracompact (UC) H\textsc{ii} region has already been detected in centimeter continuum emission (Hughes \& MacLeod 1993; Shepherd et al. 1999; Shepherd 2001), peaking at 05$^{\mbox{\scriptsize{h}}}$58$^{\mbox{\scriptsize{m}}}$13$^{\mbox{\scriptsize{s}}}$.531, 16$^{\circ}$31$^{\mbox{\scriptsize{m}}}$58$^{\mbox{\scriptsize{s}}}$.29 (J2000).
Based on the measured spectral index of 0.3$\pm$0.07  (Shepherd et al. 2004) and the 1.5\,mJy total flux at the 3.6\,cm band  (Shepherd et al. 1999), we estimate that the free-free continuum emission contributes to 3.6-6.0\,mJy at the 0.88\,mm wavelength band.
Comparing with the first contour level of the dust thermal emission at 0.88 mm in Figures \ref{fig_pol} and \ref{fig_jet}, the free-free continuum emission is negligible.

The SMA Stokes-I continuum image (Figure \ref{fig_pol}, \ref{fig_jet}) shows a $\sim$6$''$ scale source, of which the emission peak (the G192 envelope, hereafter) at 5$^{\mbox{\scriptsize{h}}}$58$^{\mbox{\scriptsize{m}}}$13$^{\mbox{\scriptsize{s}}}$.535, 16$^{\circ}$31$'$58$''$.25 (J2000) is connected with an $\sim$3$''$ scale elongation toward the southeast.
With the $\gtrsim$1$''$ angular resolution of our observations (Section \ref{chap_obs}), the peak of the 0.88\,mm Stokes-I emission cannot be distinguished from the centimeter continuum peak. 

\subsubsection{Polarized Dust Emission}

We detect significant polarized continuum emission toward the peak of the Stokes-I emission, and detect marginal polarized continuum emission at two positions located in the southeastern elongation.
The B-field directions\footnote{The magnetic field (B-field) direction should be perpendicular to the polarization direction (Lazarian 2007 and references therein).} at the peak of the Stokes-I emission are aligned in the north-south direction. 
The B-field directions at the other two detections are roughly parallel to the southeastern elongation.

\subsubsection{The CO 3-2 line}
\label{subsub_co}
\begin{figure}[h]
\hspace{-0.8cm}
\rotatebox{-90}{
\includegraphics[width=6cm]{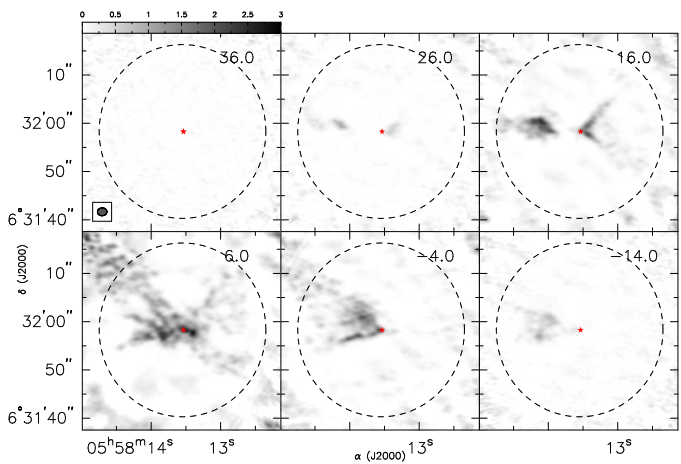}}
\caption{\footnotesize{
The velocity channel maps of CO 3-2 in the intermediate velocity range (robust 0 weighted).
The red star marks the peak of the 0.88\,mm continuum emission (R.A.: 5$^{\mbox{\scriptsize{h}}}$58$^{\mbox{\scriptsize{m}}}$13$^{\mbox{\scriptsize{s}}}$.535, Decl.: 16$^{\circ}$31$'$58$''$.25). 
Grey-scale bar has the unit of Jy\,beam$^{-1}$\,km\,s$^{-1}$.
}}
\label{fig_chan}
\end{figure}

\begin{figure}
\includegraphics[width=8.5cm]{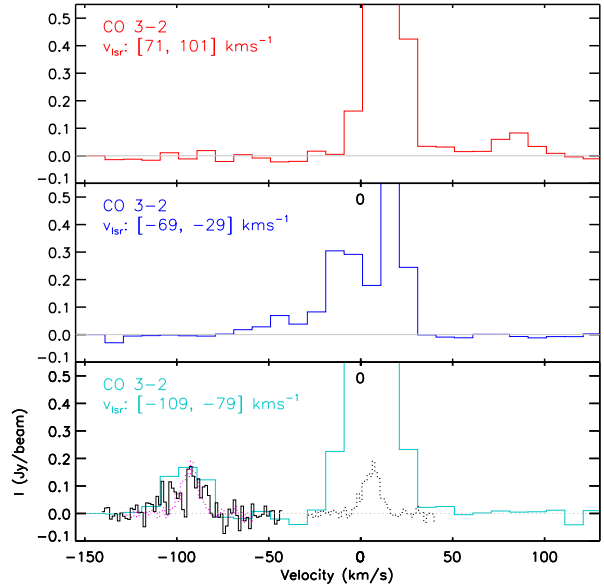}
\caption{\footnotesize{
The spectra of CO 3-2 centered at the location of the [-109,-79] km\,s$^{-1}$ component (cyan), the [-69,-29] km\,s$^{-1}$ component (blue), and the [71, 101] km\,s$^{-1}$ component as shown in Figure \ref{fig_jet}.
For each component, the spectrum is averaged in the region enclosed by the first contour levels in Figure \ref{fig_jet}.
For a comparison, we overplot the SO$_{2}$ 19$_{1,19}$-18$_{0,18}$ spectrum (scaled by a factor of 0.3) in dotted line, and overplot the manually shifted SO$_{2}$ 19$_{1,19}$-18$_{0,18}$ spectrum of which the peak is -93 km\,s$^{-1}$ (dotted magenta).
We overplot the 1.4 km\,s$^{-1}$ velocity resolution CO 3-2 spectrum for the [-109,-79] km\,s$^{-1}$ component in black solid line (1$\sigma\sim$38 mJy\,beam$^{-1}$).
}}
\label{fig_spectra}
\end{figure}

The most significant structure detected in CO 3-2 is the biconical cavity walls, centering around the peak of the Stokes-I emission (Figure \ref{fig_jet}). 
The channel maps of CO 3-2 show the east-west bipolar cavity walls (Figure \ref{fig_chan}), which is consistent with the earlier observations of CO 1-0 (Shepherd et al. 1998).
There is no zero-spacing information available and that the CO 3-2 maps for the spatially extended cavity walls cannot constrain its column density.
The CO 3-2 emission at the higher velocity ranges (e.g. $|\delta v_{lsr}|$$\gg$30\,km\,s$^{-1}$) shows redshifted (relative to the systemic velocity 5.8 km\,s$^{-1}$; see also Section \ref{sub_kinematics}) gas in west of the peak of Stokes-I continuum emission, and shows blueshifted gas around and toward the east of the peak of Stokes-I continuum emission.
The spectra of the high velocity CO gas are presented in Figure \ref{fig_spectra}.

The [$-109, -79$] km\,s$^{-1}$ component of the CO 3-2 outflow is potentially contaminated by the CH$_{3}$OH 16$_{1,15}$-15$_{2,14}$ ($\nu$: 345.90397\,GHz, E$_{u}$/k: 333\,K) and the  CH$_{3}$OH 16$_{-3,16}$-15$_{-4,14}$ ($\nu$: 345.91919\,GHz, E$_{u}$/k: 459\,K) emission from the G192 envelope. 
Assuming the averaged gas temperature in the G192 envelope to be equal to the SO$_{2}$ rotational temperature 84\,K (Section \ref{subsub_line}) and a constant CH$_{3}$OH abundance, the extrapolation of the detected CH$_{3}$OH 7$_{1,7}$-6$_{1,6}$ flux indicates that the CH$_{3}$OH 16$_{1,15}$-15$_{2,14}$ transition will contribute to an averaged integrated flux density of only $\sim$0.2 Jy\,beam$^{-1}$km\,s$^{-1}$, which is very dim as compared with the detected integrated flux density of CO in this velocity range (Figure \ref{fig_jet}). 
The CH$_{3}$OH 16$_{-3,16}$-15$_{-4,14}$ transition should be still dimmer. 
However, we cannot rule out that the abundance of CH$_{3}$OH is enhanced in the warmer center of the hot core.
The (CH$_{3}$)$_{2}$CO 15$_{8,8}$-14$_{5,9}$ EA, 15$_{8,8}$-14$_{5,9}$ AE, and 15$_{7,8}$-14$_{6,9}$ AA transitions, which were detected in Orion-KL (Friedel et al.2005), can contribute to the emission in the velocity range of $\sim$[-95.3, -93] km\,s$^{-1}$; and the $^{34}$SO$_{2}$ 17$_{4,14}$-17$_{3,15}$ transition ($\nu$: 345.92928\,GHz, E$_{u}$/k: 179\,K) can contribute to the emission at $\sim$-116 km\,s$^{-1}$.
With the observations of only one CO transition, we cannot make sure whether the [$-109, -79$] km\,s$^{-1}$ component is dominantly contributed by the CO 3-2 emission, or is very much contaminated by other molecular lines.
We will exclude this component in the following discussion.

The rest of the high velocity CO emission can support the existence of the high velocity bipolar molecular outflow in the east-west direction, emanated from the peak of Stokes-I continuum.
The powering source(s) of these high velocity gas remains to be checked by more sensitive observations. 
If the previous fittings of the inclination of the biconical cavity ($i$=63$^{\circ}$; Shepherd et al. 1998) can be applied to the $|\delta v_{lsr}|$$\gg$30\,km\,s$^{-1}$ gas, then the detected deprojected outflow velocities can be up to 210\,km\,s$^{-1}$.

\begin{table}[h]
\begin{center}
\caption{\footnotesize{The derived CO 3-2 parameters for the high velocity gas. 
}}
\label{table_outflow}
\begin{tabular}{| rl | ccc |}\hline
Component  	&		& Mass  				& Momentum 							& Energy \\
                     &   		& (10$^{-3}$$M_{\odot}$) 	& (10$^{-3}$$M_{\odot}$\,km\,s$^{-1}$) 	& (10$^{44}$ erg) \\\hline
$[$71, 101$]$	 &km\,s$^{-1}$			&	0.069		& 	10  	& 0.16	\\
$[$-69, -29$]$	 & km\,s$^{-1}$			&	1.2		& 	130  	& 1.4	\\
$[$-109, -79$]$ & km\,s$^{-1}$			&	0.28		& 	68  	& 1.7	\\\hline   
\end{tabular}
\end{center}
\footnotesize{Notes. The momentum and energy are corrected for the inclination angle of $i$=63$^{\circ}$. The $[$-109, -79$]$ km\,s$^{-1}$ component is likely to be contaminated by other molecular lines (see Section \ref{sub_geometry}), and therefore our estimates should be considered as an upper limit.
}
\vspace{0.15cm}
\end{table}

We do not have data to measure the gas temperature in the high velocity CO 3-2 outflows.
However, the $[$71, 101$]$ km\,s$^{-1}$  component is spatially close to the G192 envelope (Figure \ref{fig_jet}).
Considering the outflowing gas may be partially entrained from the G192 envelope, a comparable initial gas temperature with the G192 envelope could be a reasonable assumption, though the cooling and the shock heating are yet to be considered.
For simplicity, we assume the averaged gas temperature in all east-west high velocity CO 3-2 outflow components are comparable with that in the G192 envelope.
Table \ref{table_outflow} gives the summary of estimated energetics of the $|\delta v_{lsr}|$$\gg$30\,km\,s$^{-1}$ gas without considering the missing flux, based on the assumption of optically thin, LTE, the gas temperature of 84\,K (Section \ref{subsub_line}), [CO]/[H$_{2}$]=10$^{-4}$, and the inclination angle $i$=63$^{\circ}$ (Shepherd et al. 1998).
The energetics of the molecular outflow on the larger ($\sim$1\,pc) scale can be found in Shepherd et al. (1998). 
Dividing the separations of the $[$-69, -29$]$ km\,s$^{-1}$ and the $[$71, 101$]$ km\,s$^{-1}$ CO 3-2 emission from the G192 envelope by their averaged velocities (i.e. Momentum/Mass) implies their dynamic timescales of 910$\pm$250 years and 220$\pm$80 years, respectively. 
The momentum supply rates from these two high velocity components are 0.14$\times$10$^{-3}$ $M_{\odot}$\,km\,s$^{-1}$\,yr$^{-1}$, 0.045$\times$10$^{-3}$ $M_{\odot}$\,km\,s$^{-1}$\,yr$^{-1}$, respectively.
The summed momentum supply rate is $\gtrsim$0.19$\times$10$^{-3}$ $M_{\odot}$\,km\,s$^{-1}$\,yr$^{-1}$, which is $\sim$6.2\% of the momentum supply rate of the parsec scale CO outflow (Shepherd et al. 1998).
Besides the two water maser sources (e.g. Shepherd et al. 2004), we are not sure whether there are other (proto)stars embedded in the G192 envelope.
Deep JVLA observations to resolve the radio jet cores may allow us to see whether the high velocity molecular outflows are powered by a single dominant source.

\begin{figure}[h]
\vspace{-0.2cm}
\hspace{-0.3cm}
\begin{tabular}{ p{3.7cm} p{3.7cm} }
\includegraphics[width=5.4cm]{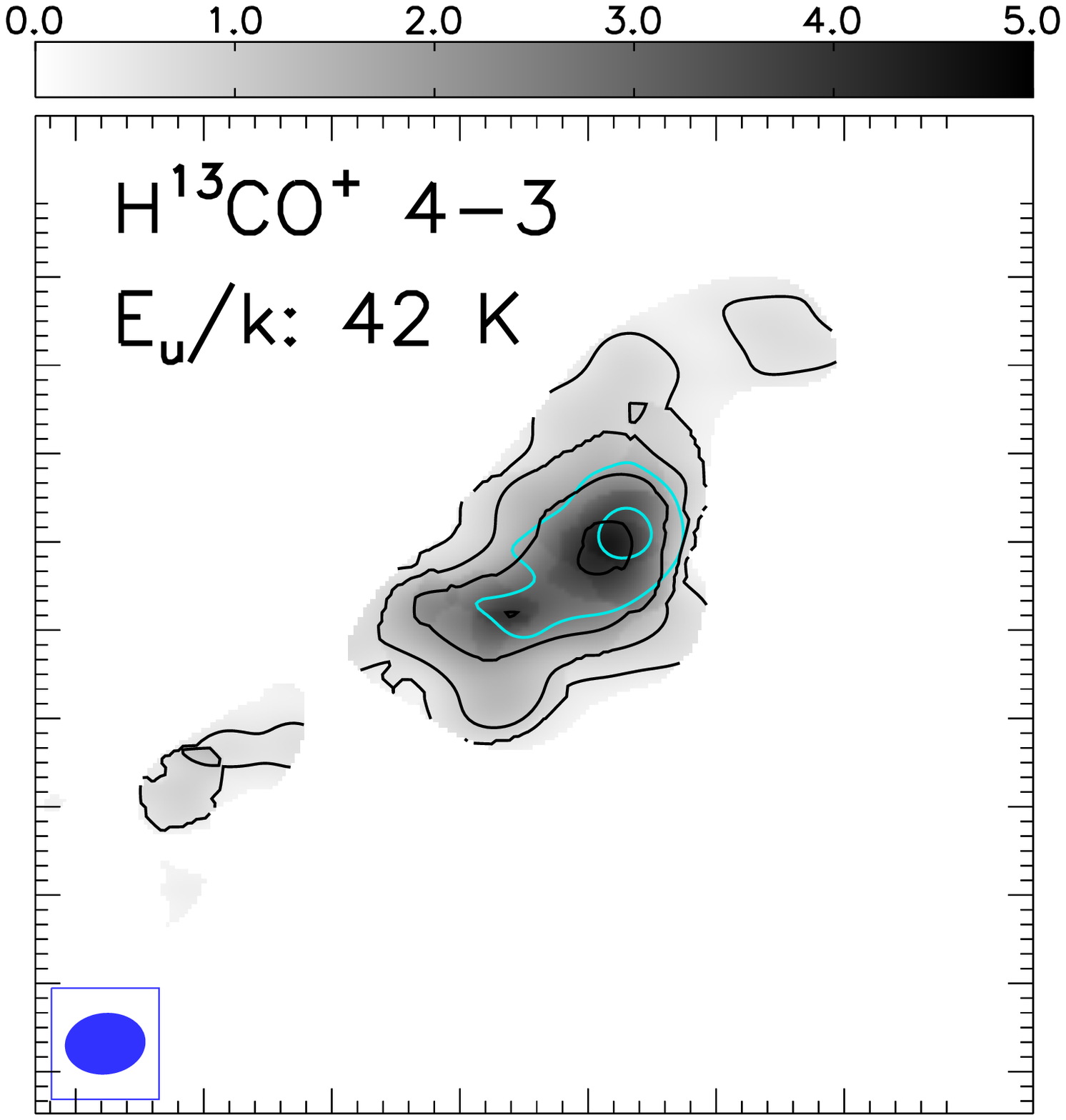} & \includegraphics[width=5.4cm]{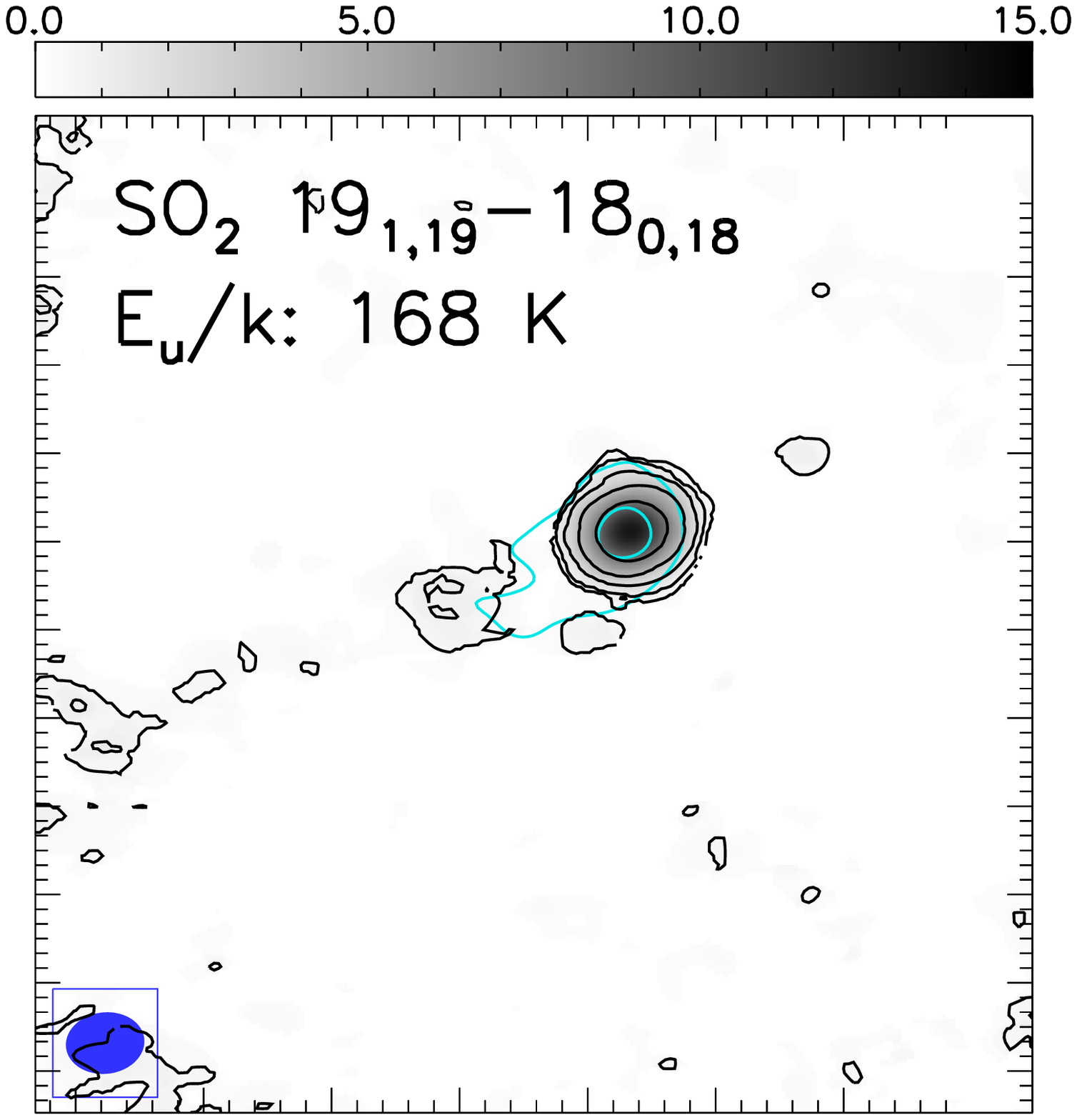}  \\
\end{tabular}

\vspace{-0.5cm}

\hspace{-0.3cm}
\begin{tabular}{ p{3.7cm} p{3.7cm} }
\includegraphics[width=5.4cm]{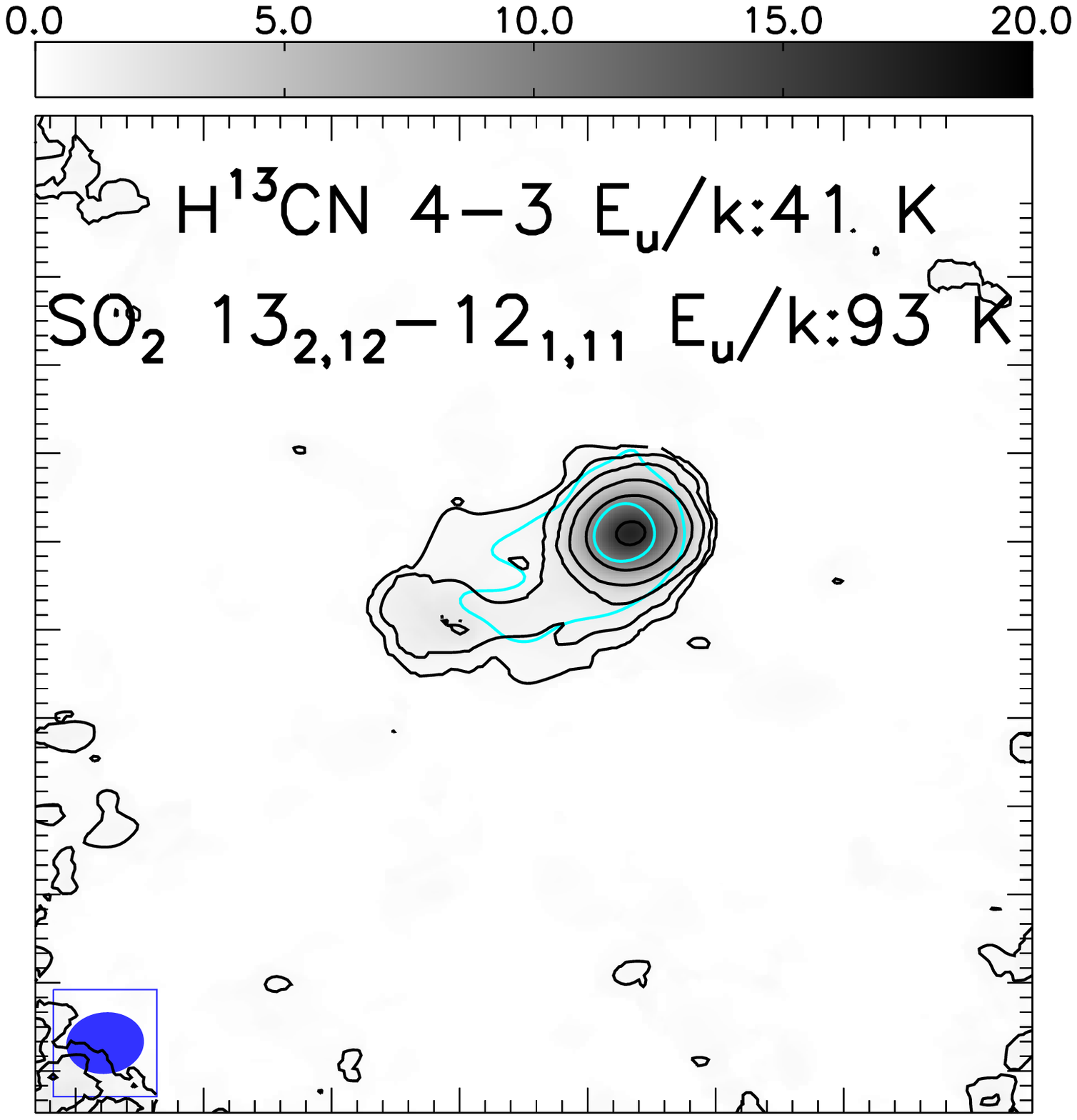} & \includegraphics[width=5.4cm]{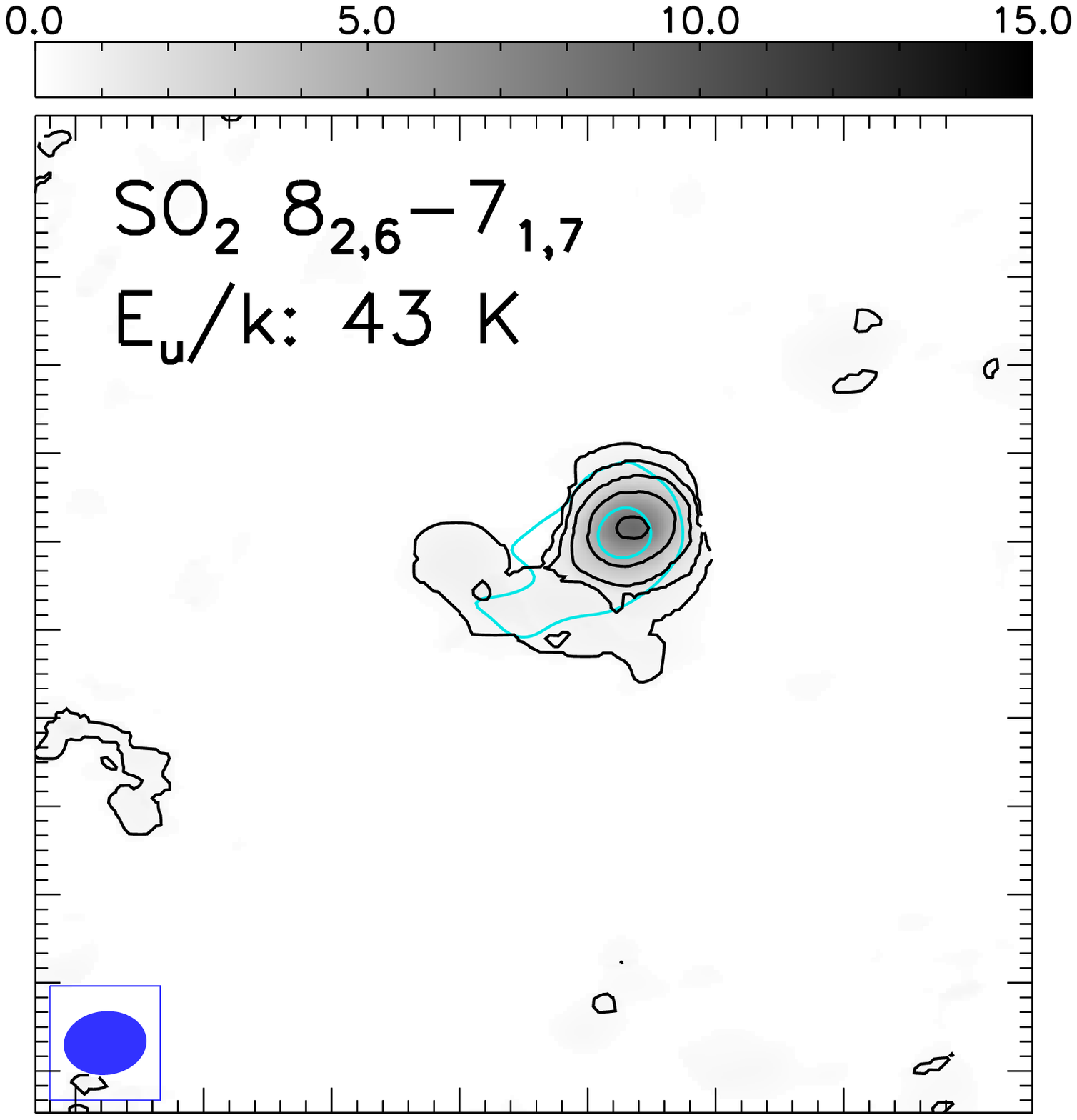}  \\
\end{tabular}

\vspace{-0.5cm}

\hspace{-0.3cm}
\begin{tabular}{ p{3.7cm} p{3.7cm} }
\includegraphics[width=5.4cm]{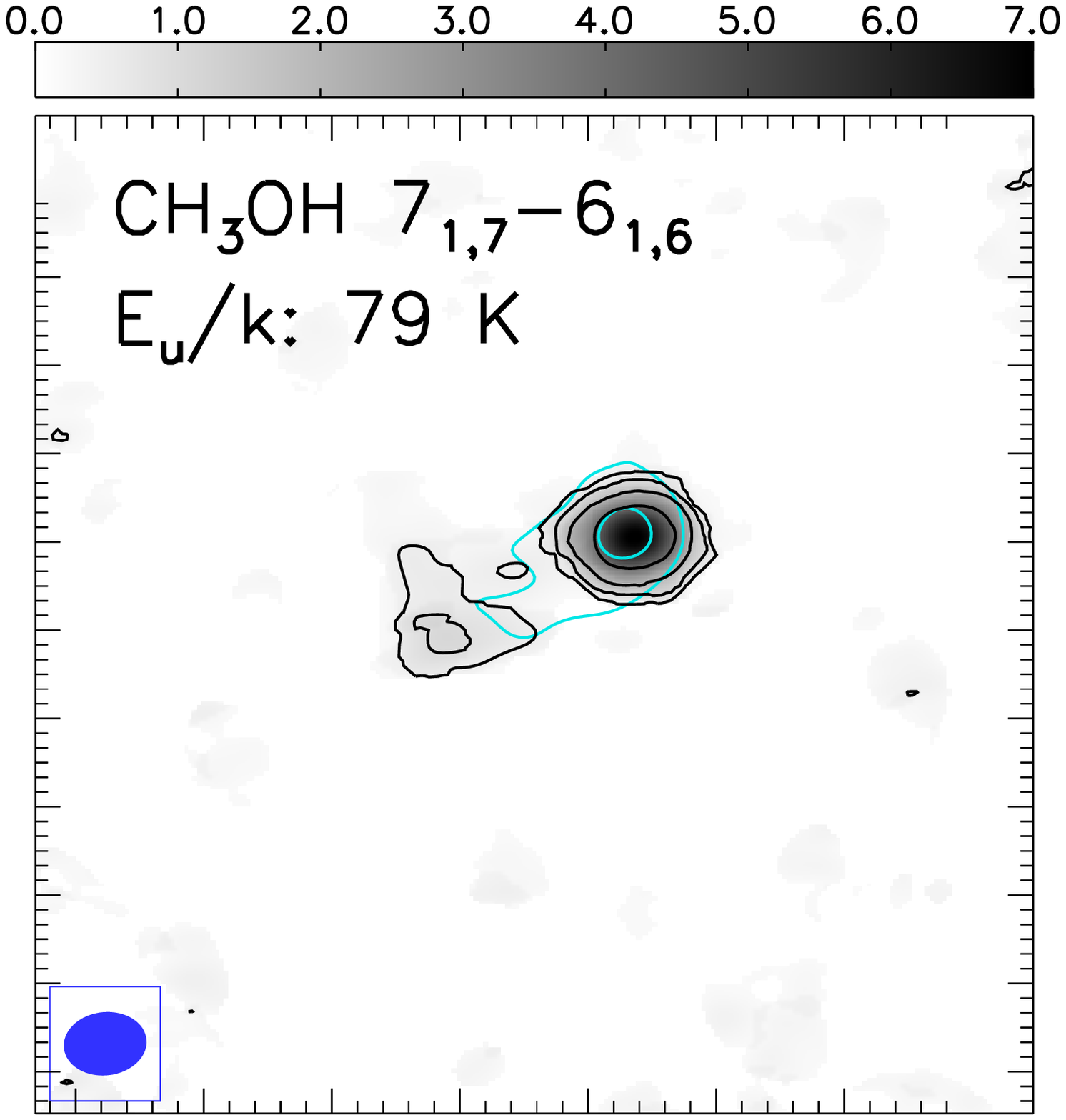} & \includegraphics[width=5.4cm]{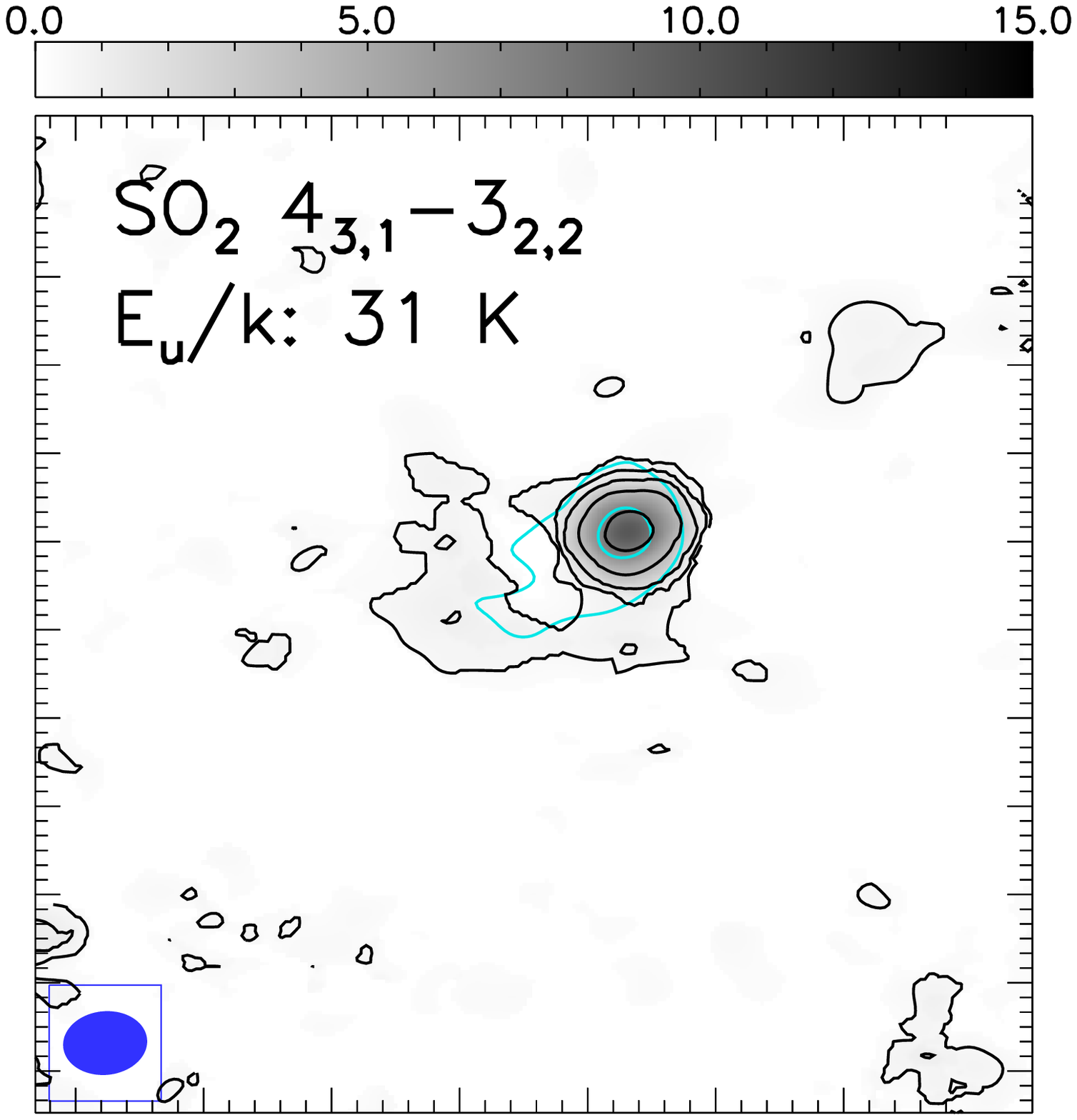}  \\
\end{tabular}

\vspace{-0.5cm}

\hspace{-0.3cm}
\begin{tabular}{ p{3.7cm} p{3.7cm} }
\includegraphics[width=5.4cm]{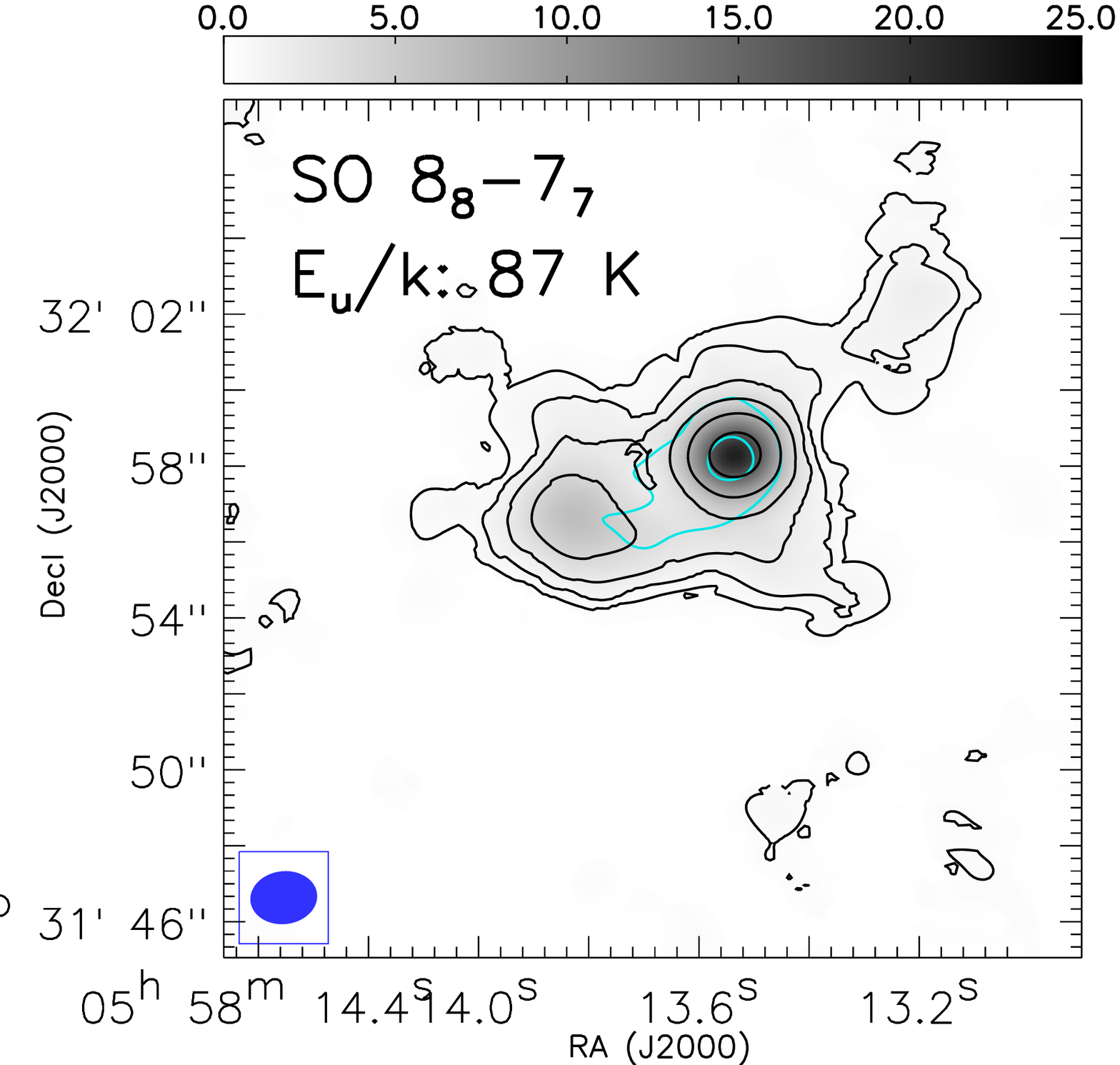} &  \\
\end{tabular}

\caption{\footnotesize{
The velocity integrated intensity maps (i.e. moment 0 maps) of the detected molecular lines (Grayscale and dark contours).
The molecular species and their upper level energy are labeled.
The H$^{13}$CN 4-3 line is blended with the hot core tracer SO$_{2}$ 13$_{2,12}$-12$_{1,11}$ (Table \ref{table_line}) and therefore shows the enhanced brightness toward the G192 envelope.
Dark contour levels are 0.5 Jy\,beam$^{-1}$\,km\,s$^{-1}$ $\times$ [1, 2, 4, 8, 16, 32, 64].
Cyan contours show the  14 mJy\,beam$^{-1}$$\times$[2, 16] levels of the 0.88\,mm continuum emission (robust 0 weighted).
Grey-scale bars have the unit of Jy\,beam$^{-1}$\,km\,s$^{-1}$.
All molecular line images in this figure are naturally weighted. 
}}
\label{fig_mnt0}
\end{figure}

\subsubsection{Other Molecular Lines}
\label{subsub_line}
The rest of the observed molecular lines trace a variety of morphology (Figure \ref{fig_mnt0}).
The observed CH$_{3}$OH transition and the SO$_{2}$ transitions mainly trace the G192 envelope, and a compact component in southeast of the G192 envelope.
The high excitation SO$_{2}$ line emission in the southeastern component may either imply a secondary source, or the interaction between the outflow and the ambient gas.
The H$^{13}$CO$^{+}$ 4-3 and H$^{13}$CN 4-3 trace both the G192 envelope and the southeastern extension. 
The SO 8$_{8}$-7$_{7}$ transition appears to trace V-shape featured by the CO redshifted gas in west of the G192 envelope, which can be associated with the inner most part of the cavity wall.
Based on the integrated flux of the SO$_{2}$ 4$_{3,1}$-3$_{2,2}$, 8$_{2,6}$-7$_{1,7}$, and 19$_{1,19}$-18$_{0,18}$ transitions in a $R$=1$''$.5 circular region centered on the G192 envelope, assuming optically thin and the local thermodynamic equilibrium (LTE), using the expression introduced in Fu et al. (2012), we constrain the rotational temperature in the G192 envelope to be T$_{rot}^{SO_{2}}$$\sim$84$^{+18}_{-13}$\,K.

\subsection{Velocity Gradient in Dense Gas}
\label{sub_kinematics}
The H$^{13}$CO$^{+}$ 4-3 line generically traces the extended dense gas (Figure \ref{fig_mnt0}, Section \ref{sub_geometry}).
Gaussian fitting of the averaged spectrum of H$^{13}$CO$^{+}$ 4-3 suggests that the systemic velocity of the dense gas is $\sim$5.8$\pm$0.4\,km\,s$^{-1}$.
The comparison of the kinematics traced by the H$^{13}$CO$^{+}$ 4-3 and the SO$_{2}$ 19$_{1,19}$-18$_{0,18}$ lines can provide clues to how the dense gas continue infalling into the inner disk/envelope.

\begin{figure}[h]
\hspace{0.5cm}
\begin{tabular}{ p{4cm} }
\includegraphics[width=8cm]{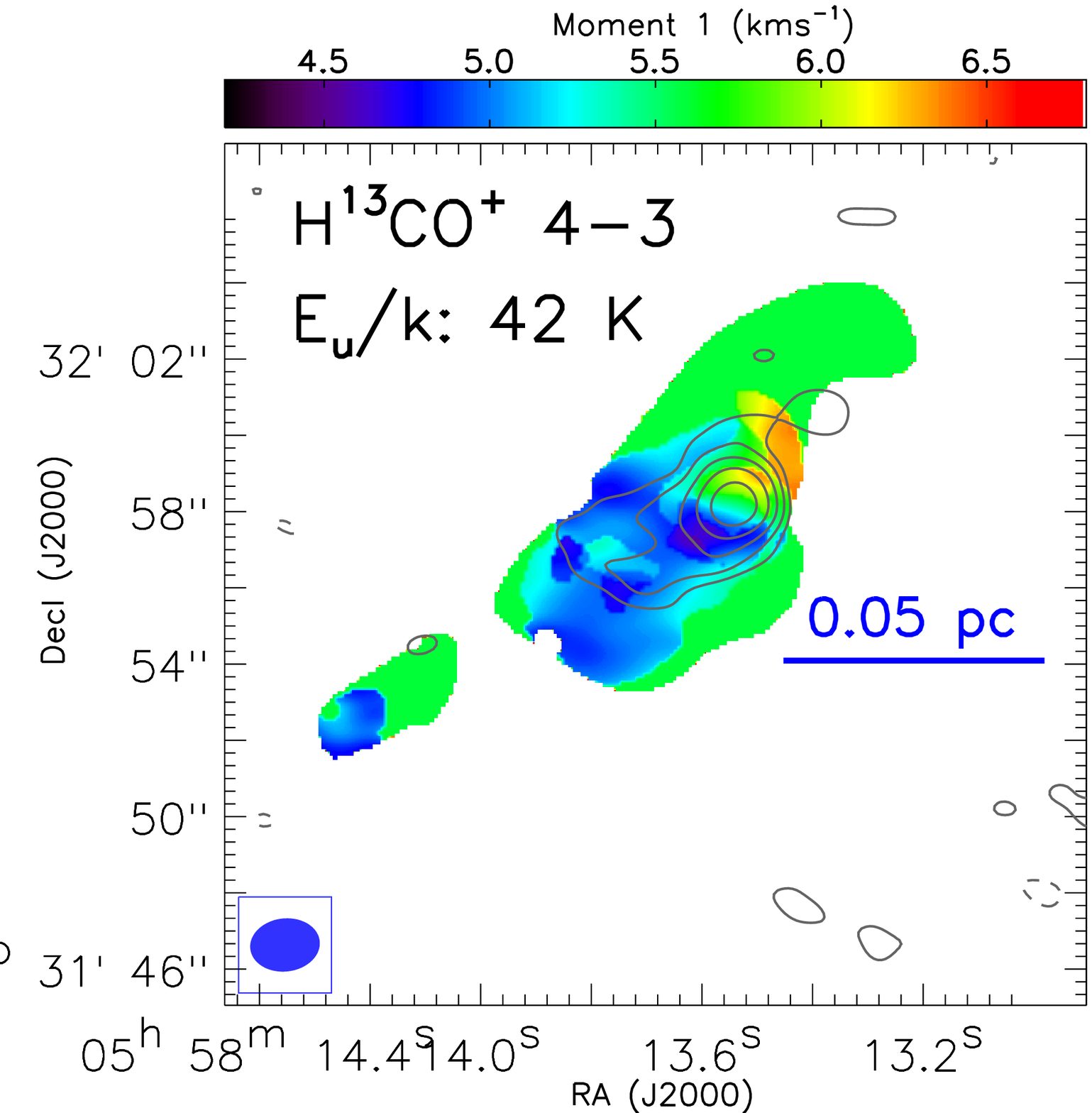} \\
\includegraphics[width=8cm]{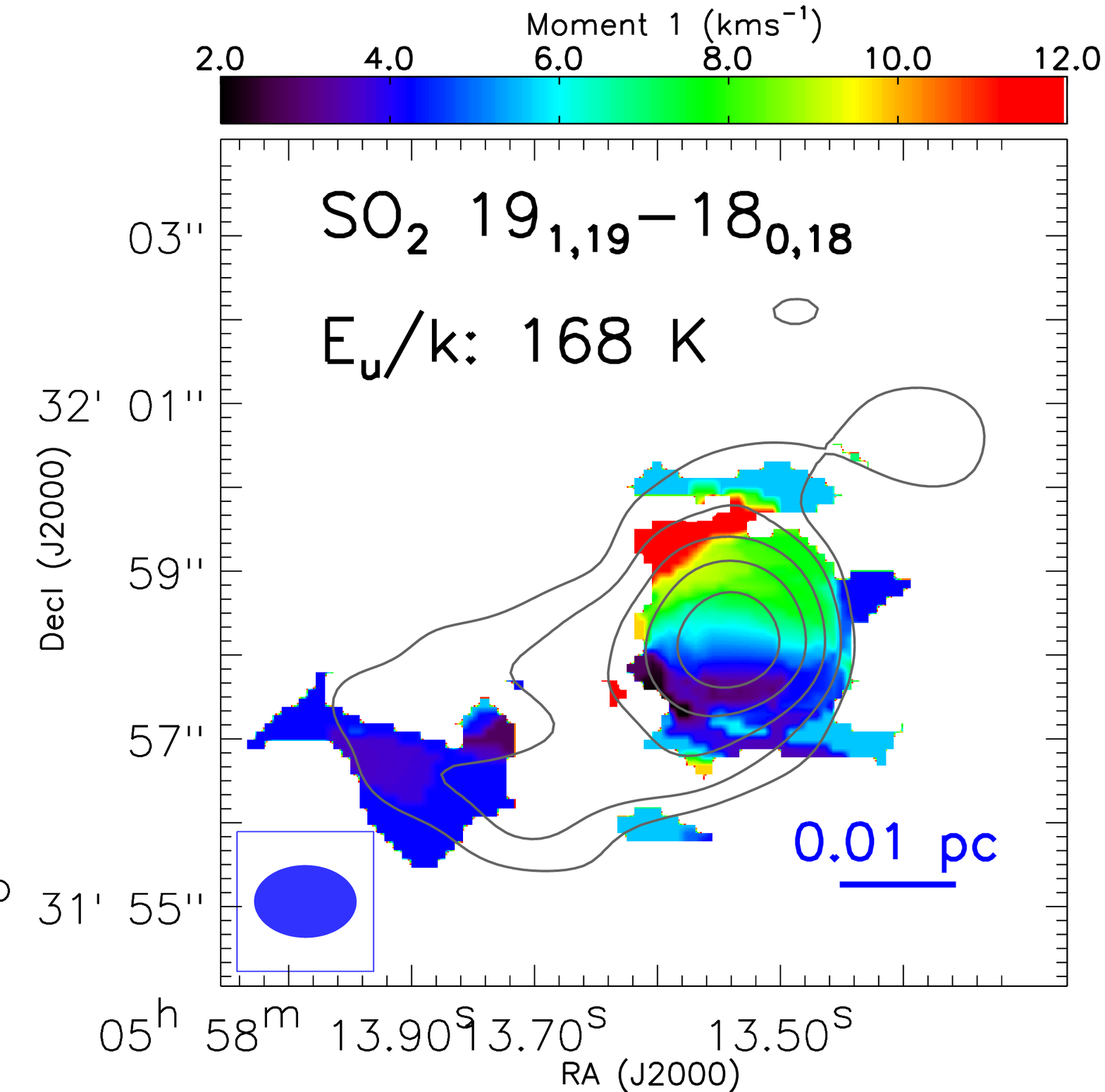} \\
\end{tabular}

\caption{\footnotesize{
The intensity-weighted average velocity maps (i.e. moment 1 maps) of SO$_{2}$ 19$_{1,19}$-18$_{0,18}$ (robust 0 weighted) and H$^{13}$CO$^{+}$ 4-3 (naturally weighted).
Gray contours show the 14 mJy\,beam$^{-1}$$\times$[-1, 1, 2, 4, 8, 16, 32] levels of the 0.88\,mm continuum emission (robust 0 weighted).
Note the scale of the upper and the bottom panels are different. 
}}
\label{fig_mnt1}
\end{figure}

Figure \ref{fig_mnt1} shows the intensity-weighted averaged velocity maps (i.e. moment 1 maps) of SO$_{2}$ 19$_{1,19}$-18$_{0,18}$ and H$^{13}$CO$^{+}$ 4-3. 
The SO$_{2}$ 19$_{1,19}$-18$_{0,18}$ transition traces a north-south velocity gradient in a $\lesssim$2$''$ region around the 0.88\,mm Stokes-I peak.
With the angular resolution of our observations, the resolved direction of the velocity gradient is consistent with the rotating plane of the disk/envelope reported by the water maser  observations (Shepherd \& Kurtz 1999; Shepherd et al. 2004; Imai et al. 2006).
In a slightly bigger area, the H$^{13}$CO$^{+}$ 4-3 line traces a northwest-southeast velocity gradient. 
The more extended gas shows the velocity close to the 5.8\,km\,s$^{-1}$ systemic velocity.
The motion discussed here is also consistently traced by the other molecular lines presented in Figure \ref{fig_mnt0}.

\begin{figure}[h]
\rotatebox{-90}{
\includegraphics[width=6cm]{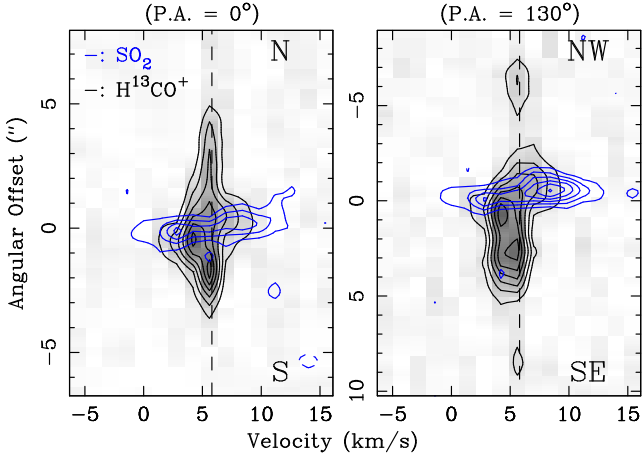}}
\caption{\footnotesize{
The position-velocity (PV) diagrams of SO$_{2}$ 19$_{1,19}$-18$_{0,18}$ (robust 0 weighted; blue contours) and H$^{13}$CO$^{+}$ 4-3 (naturally weighted; grayscale and dark contours).
The PV cuts are centered at the peak of the 0.88\,mm Stokes-I continuum emission [5$^{\mbox{\scriptsize{h}}}$58$^{\mbox{\scriptsize{m}}}$13$^{\mbox{\scriptsize{s}}}$.535, 16$^{\circ}$31$'$58$''$.25 (J2000)].
The position angles of the PV cuts are labeled in the individual panels. 
Contour levels are 0.2 Jy\,beam$^{-1}$ $\times$[1, 2, 3, 4, 5, 6].
Dashed lines label the velocity $v_{lsr}$=5.8\,km\,s$^{-1}$.
}}
\label{fig_pv}
\end{figure}

Figure \ref{fig_pv} shows the position-velocity (PV) diagrams of the H$^{13}$CO$^{+}$ 4-3 and the SO$_{2}$ 19$_{1,19}$-18$_{0,18}$ lines. 
The PV cuts are centered at the peak of the continuum emission [5$^{\mbox{\scriptsize{h}}}$58$^{\mbox{\scriptsize{m}}}$13$^{\mbox{\scriptsize{s}}}$.535, 16$^{\circ}$31$'$58$''$.25 (J2000)].
The comparison of the SO$_{2}$ 19$_{1,19}$-18$_{0,18}$ and H$^{13}$CO$^{+}$ 4-3 PV diagrams at P.A.=0$^{\circ}$ may imply an accelerated rotation of the warmer gas, which should lie closer to the embedded B3 star.
The PV diagrams at P.A.=130$^{\circ}$ show a bulk of blueshifted gas in the southeast (0$''$-5$''$).
With the 0.7\,km\,s$^{-1}$ velocity resolution of our observations, we  cannot resolve a clear trend of motion in that blueshifted gas component.

\begin{figure}[h]
\hspace{-0.3cm}
\includegraphics[width=8cm]{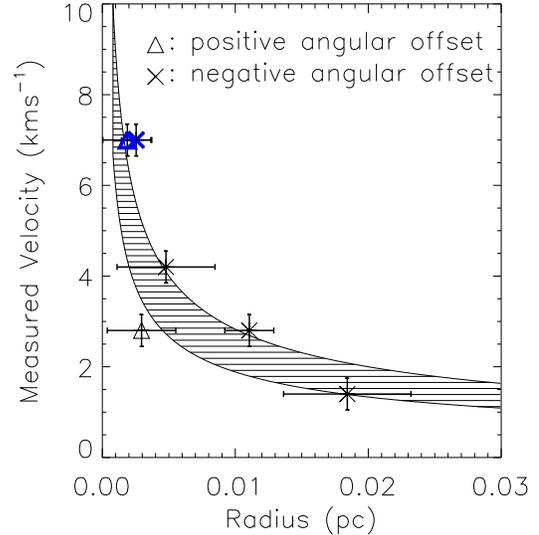}
\caption{\footnotesize{
The detected terminal velocity (relative to the $v_{lsr}$=5.8 km\,s$^{-1}$) and the projected Keplerian rotation curve. 
The origin (i.e. radius 0.0\,pc) is defined at 5$^{\mbox{\scriptsize{h}}}$58$^{\mbox{\scriptsize{m}}}$13$^{\mbox{\scriptsize{s}}}$.535, 16$^{\circ}$31$'$58$''$.25 (J2000).
Black symbols show the terminal velocities traced by the H$^{13}$CO$^{+}$ 4-3 transition.
Blue symbols show the terminal velocities traced by the SO$_{2}$ 19$_{1,19}$-18$_{0,18}$ transition.
We use different symbols to distinguish the values measured from the positive angular offset (i.e. south) and from the negative angular offset (i.e. north).
The error bars reflect the uncertainties caused by the limited 0.7\,km\,s$^{-1}$ velocity resolution.
The horizontally hatched region is bound by the two projected (inclination $i$=63$^{\circ}$) Keplerian rotation curve models, which are bound by the gravitational force of the enclosed stellar and molecular gas mass of 11.2--25.2\,M$_{\odot}$. 
}}
\label{fig_rot}
\end{figure}

We jointly analyze the velocity field presented in the PV diagrams (P.A.=0$^{\circ}$) by applying the terminal velocity method (Sofue \& Rubin 2001).
This method defines the measured terminal velocity $v_{t}$ by a velocity at which the intensity $I$ equals to 
\[I_{t} = \eta I_{max},\]
where $\eta$ is an adjustable parameter that can be optimized by checking the consistency of results from different molecular lines (e.g. Liu et al. 2010).
We consider $\eta$=0.2 is an optimal choice to analyze our data, which suppresses the confusion of gas from outer radii while guarantees that all pixels picked up by this method are above the 3$\sigma$ detection level.
We note that relative to the naive choice $\eta$=1.0, our choice of $\eta$=0.2 would potentially overestimate the rotational velocity.
Figure \ref{fig_rot} presents the $\eta$=0.2 results up to a radius that the terminal velocity still can be distinguished from the systemic velocity 5.8\,km\,s$^{-1}$.
By assuming the plane of rotation\footnote{Shepherd \& Kurtz (1999) reported that the decconvolved 2.6\,mm emission source has a position angle of -20$^{\circ}$; and the water masers lines along the axis with -44$^{\circ}$$\pm$21$^{\circ}$. The exact orientation of the rotational axis is not yet known. This uncertainty may lead to the underestimates of the radii in Figure \ref{fig_rot} by up to $\sim$30\%. This issue can be resolved by observing the proper motion of the water masers, or by observing the thermal dust continuum emission with higher angular resolution.} lies along the axis with P.A.=0$^{\circ}$, the measured terminal velocities within the 0.01 pc (1$''$.36) radius are consistent with the inclined ($i$=63$^{\circ}$) Keplerian rotation, which is bound by the gravitational force of the embedded mass $M_{*+gas+dust}\sim$11.2-25.2\,M$_{\odot}$ (Shepherd \& Kurtz 1999; Shepherd et al. 2001; Shepherd et al. 2004; Shiozaki et al. 2011).
At the $>$0.01 pc radii, the steeper decline of the measured terminal velocities than the rotational velocities of the Keplerian models may indicate that the radial infall motion is not negligible in the more extended region (see discussion in Tobin et al. 2012), although it is also consistent with the changes of the inclination angle.
The previous VLA observations of the 22\,GHz water maser further traced the [-7, 16]\,km\,s$^{-1}$ Keplerian rotation into inward of the $\sim$0$''$.25 (360 AU) radius (Shepherd \& Kurtz 1999).
These data consistently indicate the centripetally accelerated rotation toward the young star. 
Whether a Keplerian disk exists can be checked by the higher angular resolution observations.


Both Figure \ref{fig_pv} and \ref{fig_rot} present the spatial asymmetry of the brightness distribution and the velocity field in H$^{13}$CO$^{+}$ 4-3.
The spatial asymmetry of the brightness distribution is consistent with the dust continuum image (Figure \ref{fig_pol}) that also shows a southeastern extension. 
However, we cannot rule out the possibility that the observed asymmetry of the velocity field is due to the variation of the excitation condition, abundance, or is caused by the more complicated radiation transfer effects.


\begin{figure*}
\begin{center}
\includegraphics[width=13cm]{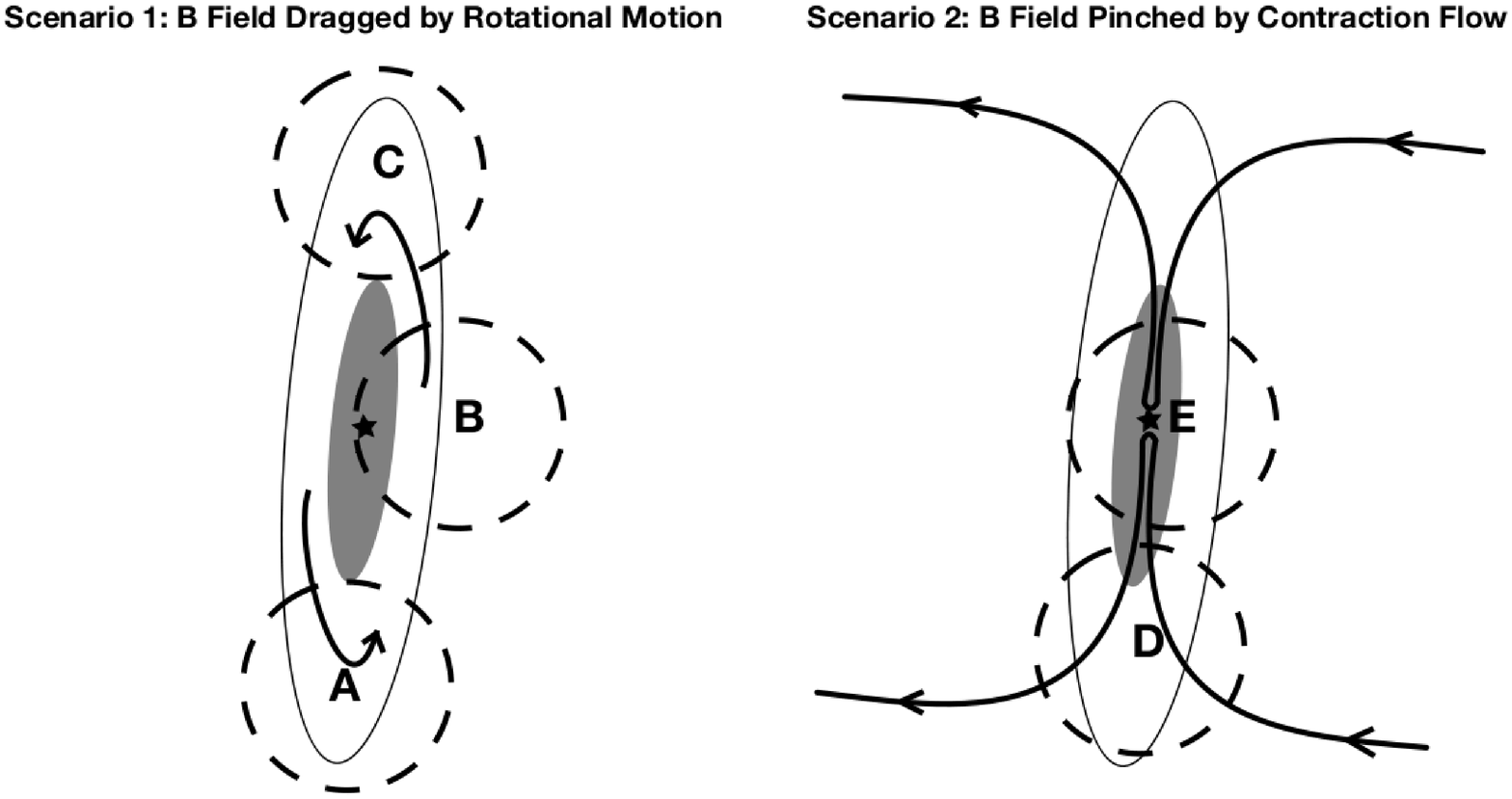}
\end{center}
\vspace{-0.6cm}
\caption{\footnotesize{The schematic pictures illustrating the observed B field (see also Section \ref{chap_summary}). The shaded areas indicate the inclined thin disk or pseudo disk around the central stars. The B field lines are shown by the thick curves with arrows. The dashed circles labeled with A-E are the observational (synthesized) beams. In beam A and C, the B field lines are nearly parallel to the line-of-sight, thus the dust emission is only weakly polarized. In beam D, the observed polarization percentage can be low because of canceling. In beam B and E, we can detect the B field aligned in the disk plane.}}
\label{fig_schematic}
\end{figure*}

\section{Discussion}
\label{chap_summary}
We observed the 0.88\,mm polarized thermal dust emission and the molecular line emission which trace a range of excitation conditions,  toward the B3 star-forming region G192.16$-$3.84.
The kinematics traced by the observed molecular lines as well as the previous detections of the 22\,GHz water maser suggest a Keplerian motion of dense gas, continuing from the $<$0.02\,pc radius to inward of the $\le$290 AU radius.
The magnetic field lines at 0.02-0.05 pc scale may have already been twisted by the rotational motion, such that the B-field directions are parallel to the plane of rotation.
The high velocity molecular outflow is observed to be perpendicular to the rotational plane of the dense gas.
Supposedly the high velocity molecular outflow is driven by the magnetocentrifugal wind, over the past $\gtrsim$10$^{3}$ years outflow dynamical timescale. 
The magnetic field lines at sub-AU scale may still be organized in the way such that they can drive the east-west bipolar wind.
We refer to Hull et al. (2012) for the discussion about the misalignment of magnetic fields and outflows in lower mass protostellar cores. 

The lower polarization fraction at the Stokes-I continuum peak (Figure \ref{fig_pol}) can be explained by the blended magnetic field lines in the inner and outer radii (Tang et al. 2009; Frau et al.2011; Padovani et al. 2012), though such an interpretation is not unique.
This "de-polarization effect" can alternatively be explained by the difference in dust grain properties or grain alignment efficiency (Lazarian 2007), although the significance of this effect is not yet robustly argued by the SMA case studies.
We tentatively explain the non detection of the polarized intensity $\sim$1$''$ northwest and southeast of the Stokes-I continuum peak by the nearly parallel to line-of-sight orientation of the magnetic field lines dragged by the rotational velocity field. 
We note that the B-field morphology is essentially unresolved.  In an edge-on geometry, if the B field is pulled in by the contraction process, with an hour-glass morphology  (e.g. Greaves et al. 1994; Girart et al. 2006, 2009; Rao et al. 2009), most of the unresolved B field lines will be aligned along the line-of-sight, or blended and cancelled across the line of sight.  The only remaining field detectable in such a case, might be the field lines which exit from the top and bottom of the edge-on disk, and hence aligned parallel to the disk.  This would be an alternative to the scenario where the field lines are rendered toroidal by rotational motions.
Figure \ref{fig_schematic} shows the schematic pictures for the proposed scenarios.

Our observations indicate that the B3 (proto)star embedded in G192.16$-$3.84 can form via a process similar to the scaled-up solar-mass type star formation.
With the sensitivity of the current instruments, observation with a large number of (fainter) samples is not yet possible.
The proposed scenario is therefore subjected to the concern of the target selection bias. 
Furthermore, in the environments of the more crowded OB clusters, whether the sources like G192.16$-$3.84 are representative, remains an open question.

\acknowledgments
The SMA data were taken as part of the Large SMA Dust Polarization Survey (PI: Qizhou Zhang).
We acknowledge the supports from the SMA staffs.
HBL thanks Vivien H.-R. Chen for useful discussions. 
JMG is supported by the Spanish MINECO AYA2011-30228-C03-02 and the Catalan AGAUR  2009SGR1172 grants. 

{\it Facilities:} \facility{SMA}










\begin{thebibliography}{}


\bibitem[Cesaroni et al.(2005)]{2005A&A...434.1039C} Cesaroni, R., Neri, R., Olmi, L., et al.\ 2005, \aap, 434, 1039 

\bibitem[Codella et al.(1996)]{1996A&A...311..971C} Codella, C., Felli, M., \& Natale, V.\ 1996, \aap, 311, 971 

\bibitem[Curran \& Chrysostomou(2007)]{2007MNRAS.382..699C} Curran, R.~L., \& Chrysostomou, A.\ 2007, \mnras, 382, 699 

\bibitem[Devine et al.(1999)]{1999AJ....117.2919D} Devine, D., Bally, J., Reipurth, B., Shepherd, D., \& Watson, A.\ 1999, \aj, 117, 2919 

\bibitem[Frau et al.(2011)]{2011A&A...535A..44F} Frau, P., Galli, D., \& Girart, J.~M.\ 2011, \aap, 535, A44

\bibitem[Friedel et al.(2005)]{2005ApJ...632L..95F} Friedel, D.~N., Snyder, L.~E., Remijan, A.~J., \& Turner, B.~E.\ 2005, \apjl, 632, L95

\bibitem[Fu et al.(2012)]{2012ApJ...746...42F} Fu, R.~R., Moullet, A., Patel, N.~A., et al.\ 2012, \apj, 746, 42

\bibitem[Girart et al.(2006)]{2006Sci...313..812G} Girart, J.~M., Rao, R., \& Marrone, D.~P.\ 2006, Science, 313, 812

\bibitem[Girart et al.(2009)]{2009Sci...324.1408G} Girart, J.~M., Beltr{\'a}n, M.~T., Zhang, Q., Rao, R., \& Estalella, R.\ 2009, Science, 324, 1408

\bibitem[Greaves et al.(1994)]{1994A&A...284L..19G} Greaves, J.~S., Murray, A.~G., \& Holland, W.~S.\ 1994, \aap, 284, L19

\bibitem[Hodapp(1994)]{1994ApJS...94..615H} Hodapp, K.-W.\ 1994, \apjs, 94, 615 

\bibitem[Hildebrand et al.(2000)]{2000PASP..112.1215H} Hildebrand, R.~H., Davidson, J.~A., Dotson, J.~L., et al.\ 2000, \pasp, 112, 1215


\bibitem[Ho et al.(2004)]{2004ApJ...616L...1H} Ho, P.~T.~P., Moran, J.~M., \& Lo, K.~Y.\ 2004, \apjl, 616, L1 


\bibitem[Hughes \& MacLeod(1993)]{1993AJ....105.1495H} Hughes, V.~A., \& MacLeod, G.~C.\ 1993, \aj, 105, 1495

\bibitem[Hull et al.(2013)]{2013ApJ...768..159H} Hull, C.~L.~H., Plambeck, R.~L., Bolatto, A.~D., et al.\ 2013, \apj, 768, 159

\bibitem[Imai et al.(2006)]{2006PASJ...58..883I} Imai, H., Omodaka, T., Hirota, T., et al.\ 2006, \pasj, 58, 883

\bibitem[Indebetouw et al.(2003)]{2003ApJ...596L..83I} Indebetouw, R., Watson, C., Johnson, K.~E., Whitney, B., \& Churchwell, E.\ 2003, \apjl, 596, L83 

\bibitem[Keto \& Zhang(2010)]{2010MNRAS.406..102K} Keto, E., \& Zhang, Q.\ 2010, \mnras, 406, 102







\bibitem[Lazarian(2007)]{2007JQSRT.106..225L} Lazarian, A.\ 2007, \jqsrt, 106, 225 



\bibitem[Liu et al.(2010)]{2010ApJ...722..262L} Liu, H. B., Ho,  P.~T.~P., Zhang, Q. et al.\ 2010, \apj, 722, 262


\bibitem[Marrone et al.(2006)]{2006PhD} Marrone, D. P. 2006 \textit{PhD thesis} Harvard University

\bibitem[Molinari et al.(1996)]{1996A&A...308..573M} Molinari, S., Brand, J., Cesaroni, R., \& Palla, F.\ 1996, \aap, 308, 573 

\bibitem[Padovani et al.(2012)]{2012A&A...543A..16P} Padovani, M., Brinch, C., Girart, J.~M., et al.\ 2012, \aap, 543, A16

\bibitem[Qiu et al.(2008)]{2008ApJ...685.1005Q} Qiu, K., Zhang, Q., Megeath, S.~T., et al.\ 2008, \apj, 685, 1005 



\bibitem[Rao et al.(2009)]{2009ApJ...707..921R} Rao, R., Girart, J.~M., Marrone, D.~P., Lai, S.-P., \& Schnee, S.\ 2009, \apj, 707, 921

\bibitem[Shepherd \& Churchwell(1996)]{1996ApJ...472..225S} Shepherd, D.~S., \& Churchwell, E.\ 1996, \apj, 472, 225 

\bibitem[Shepherd et al.(1998)]{1998ApJ...507..861S} Shepherd, D.~S., Watson, A.~M., Sargent, A.~I., \& Churchwell, E.\ 1998, \apj, 507, 861

\bibitem[Shepherd \& Kurtz(1999)]{1999ApJ...523..690S} Shepherd, D.~S., \& Kurtz, S.~E.\ 1999, \apj, 523, 690 

\bibitem[Shepherd et al.(2001)]{2001Sci...292.1513S} Shepherd, D.~S., Claussen, M.~J., \& Kurtz, S.~E.\ 2001, Science, 292, 1513

\bibitem[Shepherd et al.(2004)]{2004ApJ...614..211S} Shepherd, D.~S., Borders, T., Claussen, M., Shirley, Y., \& Kurtz, S.\ 2004, \apj, 614, 211

\bibitem[Shiozaki et al.(2011)]{2011PASJ...63.1219S} Shiozaki, S., Imai, H., Tafoya, D., et al.\ 2011, \pasj, 63, 1219

\bibitem[Sridharan et al.(2005)]{2005ApJ...631L..73S} Sridharan, T.~K., Williams, S.~J., \& Fuller, G.~A.\ 2005, \apjl, 631, L73 

\bibitem[Shu et al.(1987)]{1987ARA&A..25...23S} Shu, F.~H., Adams, F.~C., \& Lizano, S.\ 1987, \araa, 25, 23 

\bibitem[Sofue \& Rubin(2001)]{2001ARA&A..39..137S} Sofue, Y., \& Rubin, V.\ 2001, \araa, 39, 137

\bibitem[Snell et al.(1990)]{1990ApJ...352..139S} Snell, R.~L., Dickman, R.~L., \& Huang, Y.-L.\ 1990, \apj, 352, 139

\bibitem[Su et al.(2007)]{2007ApJ...671..571S} Su, Y.-N., Liu, S.-Y., Chen, H.-R., Zhang, Q., \& Cesaroni, R.\ 2007, \apj, 671, 571

\bibitem[Tang et al.(2009)]{2009ApJ...700..251T} Tang, Y.-W., Ho, P.~T.~P., Koch, P.~M., et al.\ 2009, \apj, 700, 251

\bibitem[Tobin et al.(2012)]{2012ApJ...748...16T} Tobin, J.~J., Hartmann, L., Bergin, E., et al.\ 2012, \apj, 748, 16 

\bibitem[Wilner \& Welch(1994)]{1994ApJ...427..898W} Wilner, D.~J., \& Welch, W.~J.\ 1994, \apj, 427, 898

\bibitem[Zhang et al.(1998)]{1998ApJ...505L.151Z} Zhang, Q., Hunter, T.~R., \& Sridharan, T.~K.\ 1998, \apjl, 505, L151

\end{thebibliography}
\end{document}